\documentclass[a4paper]{article}

\usepackage[utf8]{inputenc}
\usepackage[nocompress]{cite}
\usepackage{fullpage}
\usepackage{hyperref}
\usepackage{amsmath, amssymb}
\usepackage{graphicx}
\usepackage{xcolor}
\usepackage{enumitem}
\newtheorem{theorem}{Theorem}

\newtheorem{lemma}{Lemma}

\newcommand{\Predecessor}{\ensuremath{\mathsf{predecessor}}}
\newcommand{\Insert}{\ensuremath{\mathsf{insert}}}
\newcommand{\Delete}{\ensuremath{\mathsf{delete}}}
\newcommand{\ParallelMember}{\ensuremath{\mathsf{pMember}}}
\newcommand{\Member}{\ensuremath{\mathsf{member}}}
\newcommand{\Compress}{\ensuremath{\mathsf{compress}}}
\newcommand{\Spread}{\ensuremath{\mathsf{spread}}}
\newcommand{\Retrieve}{\ensuremath{\mathsf{pRetrieve}}}

\newcommand{\lab}{\ensuremath{\mathsf{label}}}
\newcommand{\str}{\ensuremath{\mathsf{str}}}
\newcommand{\key}{\ensuremath{\mathsf{key}}}
\newcommand{\minl}{\ensuremath{\mathsf{min}}}
\newcommand{\maxl}{\ensuremath{\mathsf{max}}}
\newcommand{\addr}[1]{\ensuremath{\mathsf{addr}(#1)}}
\newcommand{\data}[1]{\ensuremath{\mathsf{data}(#1)}}

\renewcommand{\angle}[1]{\langle{#1}\rangle}

\newcommand{\uw}[1]{\ensuremath{#1}}
\newcommand{\uwi}[2]{\ensuremath{#1\langle #2 \rangle}}

\title{Predecessor on the Ultra-Wide Word RAM\thanks{An extended abstract appeared at the \emph{18th Scandinavian Symposium and Workshops on Algorithm Theory}~\cite{BGS2022b}.}}
\author{Philip Bille\footnote{Supported by Danish Research Council grant DFF-8021-002498.} \and Inge Li G{\o}rtz$^\dagger$ \and Tord Stordalen }
\date{}

\begin{document}
\maketitle
\begin{centering}
\large 
Technical University of Denmark, DTU Compute, Kgs.~Lyngby, Denmark

\texttt{\{phbi,inge,tjost\}@dtu.dk}

\end{centering}

\begin{abstract}
    We consider the predecessor problem on the ultra-wide word RAM model of computation, which extends the word RAM model with \emph{ultrawords} consisting of $w^2$ bits [TAMC, 2015]. The model supports arithmetic and boolean operations on ultrawords, in addition to \emph{scattered} memory operations that access or modify $w$ (potentially non-contiguous) memory addresses simultaneously. The ultra-wide word RAM model captures (and idealizes) modern vector processor architectures.
    
    Our main result is a simple, linear space data structure that supports predecessor in constant time and updates in amortized, expected constant time. This improves the space of the previous constant time solution that uses space in the order of the size of the universe. Our result holds even in a weaker model where ultrawords consist of $w^{1+\epsilon}$ bits for any $\epsilon > 0 $. It is based on a new implementation of the classic $x$-fast trie data structure of Willard [Inform.~Process.~Lett. 17(2), 1983] combined with a new dictionary data structure that supports fast parallel lookups. 
\end{abstract}

\section{Introduction}

Let $S$ be a set of $n$ $w$-bit integers. The \emph{predecessor problem} is  to maintain $S$ under the following operations.
    \begin{itemize}
        \item $\Predecessor(x)$: return the largest $y \in S$ such that $y \leq x$.
        \item $\Insert(x)$: add $x$ to $S$.
        \item $\Delete(x)$: remove $x$ from $S$.
    \end{itemize}
The predecessor problem is a fundamental and well-studied data structure problem, both from the perspective of upper bounds~\cite{PT2006, PT2014, BKZ1997, Willard1983, FW1993, Andersson1996, BF2002, Boas1977, BBV2010, BBPV2009} and lower bounds~\cite{Ajtai1988, BF2002, Miltersen1994, MNSW1998, SV2008, PT2006, PT2007}. The problem has many applications, for instance integer sorting~\cite{Andersson1996, AHNR1998, FW1993, Han2004}, string sorting~\cite{AFGV1997, BFK2006, Farach1997}, and string searching~\cite{Belazzougui2012, BBV2010, BEGV2018, BGS2017, BLRS+2015}. See Navarro and Rojas-Ledesma~\cite{NR2020} for a recent survey. 

On the word RAM model of computation, the complexity of the problem is well-understood with the following tight upper and lower bound on the time for operations given by P\u{a}tra\c{s}cu and Thorup~\cite{PT2014}.  
\begin{equation}    
  \Theta\left(\max \left[1, \min \left\{ 
         \log_w n,
         \frac{\log \frac{w}{\log w}}{\log \left( \log \frac{w}{\log w} / \log \frac{\log n}{\log w}\right)},
         \log \frac{\log(2^w - n)}{\log w}
\right\}\right]\right).
\end{equation}
From the upper bound perspective, the first branch matches dynamic fusion trees~\cite{PT2014}, the second branch is based on an extension of the techniques from Beame and Fich~\cite{BF2002}, and the last branch is based on an extension of dynamic van Emde Boas trees~\cite{BKZ1997}. Note that the lower bound implies that we cannot support operations in constant time for general $n$ and $w$. Hence, a natural question is if practical models of computation capturing modern hardware can allow us to overcome the superconstant lower bound.  

One such model is the \emph{RAM with byte overlap} (RAMBO) by Brodnik et al.~\cite{BCFKM2005}. This model extends the word RAM model by adding a set of special words that share bits; flipping a bit in one word will also affect all the other words that share that bit. The precise model is determined by the layout of the shared bits. It is feasible to make hardware based on this model, and prototypes have been built~\cite{LMSTBK1999}. In the RAMBO model, Brodnik et al.~\cite{BCFKM2005} gave a predecessor data structure using constant time per operation with $O(2^w/w)$ space (counting both regular words and shared words). They also gave a randomized version of the solution that uses constant time with high probability and reduces the regular space to $O(n)$ (but still needs $\Omega(2^w/w)$ space for the shared words). In both cases, the total space is near-linear in the size of the universe.

More recently, Farzan et al.~\cite{FLNS2015} introduced the \emph{ultra-wide word RAM model} (UWRAM). The UWRAM extends the word RAM model by adding special \emph{ultrawords} of $w^2$ bits. The model supports standard boolean and arithmetic operations on ultrawords, as well as \emph{scattered} memory operations that access $w$ words in memory in parallel. The UWRAM model captures (and idealizes) modern vector processing  architectures~\cite{Reinders2013, SBBE+2017, CRDI2007} (see Section~\ref{sec:uwram_model} for details of the model). Farzan et al.~\cite{FLNS2015} showed how to simulate algorithms for the RAMBO model on the UWRAM at the cost of increasing the space by a polylogarithmic factor. Simulating the above RAMBO solution for the predecessor problem, they gave a solution to the predecessor problem on the UWRAM using worst case constant time for all operations and $O(w2^w)$ space. 

\subsection{Our Results}
We revisit the predecessor problem on the UWRAM and show the following main result.
\begin{theorem}\label{thm:constant_time_predecessor}
    Given a set of $n$ $w$-bit integers, we can construct an $O(n)$ space data structure on a UWRAM that supports $\Predecessor$ in constant time and $\Insert$ and $\Delete$ in amortized expected constant time. The result holds even when ultrawords consist of $w^{1+\epsilon}$ bits for any fixed $\epsilon > 0$.
\end{theorem}
Compared to the previous result of Farzan et al.~\cite{FLNS2015}, Theorem~\ref{thm:constant_time_predecessor} significantly reduces the space from $O(w2^w)$ to linear  while maintaining constant time for operations (note that query time is worst-case, while updates are amortized expected). Furthermore, our result works in a weaker model were ultrawords consist of only $w^{1+\epsilon}$ bits for any arbitrarily small $\epsilon > 0$. In this restricted model we limit our reliance on the powerful scattered memory operations by allowing them to access only $w^\epsilon$ words in memory in parallel.

A key component in our solution is a new dictionary data structure of independent interest that supports fast parallel lookups on the UWRAM. We define the problem as follows. Recall that an ultraword $X$ consists of $w^2$ (or $w^{1+\epsilon}$) bits. We view $X$ as divided into $w$ (or $w^{\epsilon}$) words of $w$ consecutive bits each, numbered from right to left starting from $0$. The $i$th word in $X$ is denoted $X\langle i \rangle$ (we discuss the model in detail in Section~\ref{sec:uwram_model}).  Given a set $S$ of $n$ $w$-bit integers, the \emph{$w^\epsilon$-parallel dictionary problem} is to maintain $S$ under the following operations.
   \begin{itemize}
        \item $\ParallelMember(\uw{X})$: return an ultraword $\uw{I}$ where $\uwi{I}{i} = 1$ if $\uwi{X}{i} \in S$ and $\uwi{I}{i} = 0$ otherwise. 
        \item $\Insert(x)$: Add $x$ to $S$.
        \item $\Delete(x)$: Remove $x$ from $S$.
    \end{itemize}
Thus, $\ParallelMember$ takes an ultraword $X$ of $w^\epsilon$ integers and returns an ultraword encoding which of these integers are in $S$. To the best of our knowledge, the $w^\epsilon$-parallel dictionary problem has not been studied before. We show the following result.
\begin{theorem}\label{thm:parallel_dictionaries}
    Given a set of $n$ $w$-bit integers on a UWRAM with $w^{1+\epsilon}$-bit ultrawords for any fixed $\epsilon > 0$, we can construct an $O(n + w^\epsilon)$-space data structure that supports $\ParallelMember$ queries in worst case constant time and $\Insert$ and $\Delete$ in amortized expected constant time.
\end{theorem}
Note that the queries are worst-case constant time, while the updates are amortized expected constant time. The time bounds of Theorem~\ref{thm:parallel_dictionaries} thus match the well-known dynamic perfect hashing structure of Dietzfelbinger et al.~\cite{DKMHRT1994} (which is also the basis of our solution), except that the queries are parallel. The space is linear except for the additive $w^\epsilon$ term, which is needed even for storing the input to the $\ParallelMember$ query.

\subsection{Techniques}\label{sec:techniques}
Our results are achieved by novel and efficient parallel implementations of well-known sequential data structures.

Our parallel dictionary structure of Theorem~\ref{thm:parallel_dictionaries} is based on the dynamic perfect hashing structure of Dietzfelbinger et al.~\cite{DKMHRT1994}. This is a two-level data structure similar to the classic static perfect hashing structure of Fredman et al.~\cite{FKS1984}. At the first level, a universal hash function partitions the input into smaller subsets, each of which is then resolved at the second level using another universal hash function mapping the elements into sufficiently large tables. The structure supports (sequential) membership queries in worst-case constant time by evaluating the hash functions and navigating the structure accordingly. Updates are supported in amortized expected constant time by carefully rebuilding and rehashing the structure during execution. At any point in time the structure never uses more than $O(n)$ space. We show how to parallelize the evaluation of a universal hash function (the simple and practically efficient \emph{multiply-shift} hash function). Then, using the scattered memory access operations, we show how to access the corresponding entries in  the structure in parallel. 
Our technique requires only small changes to the structure of Dietzfelbinger et al.~\cite{DKMHRT1994} and we can directly apply their update operations to our solution. Thus, we are able to parallelize the worst-case constant time sequential membership query while maintaining the amortized expected constant update time bound of Dietzfelbinger et al.~\cite{DKMHRT1994}, leading to the bounds of Theorem~\ref{thm:parallel_dictionaries}.

We first show Theorem~\ref{thm:constant_time_predecessor} for the simpler case $\epsilon = 1$ that corresponds to the original UWRAM model by~\cite{FLNS2015}. Our data structure is based on the emph{$x$-fast trie} of Willard~\cite{Willard1983} combined with our parallel dictionary structure of Theorem~\ref{thm:parallel_dictionaries}. The $x$-fast trie consists of the trie $T$ of the binary representation of the input set. Also, at each level $i$, the structure stores a dictionary containing the length-$i$ prefixes of the input set. In total, this uses $O(nw)$ space. The $x$-fast trie supports predecessor queries in $O(\log w)$ time by binary searching the levels (with the help of the dictionaries) to find the longest common prefix of the query and the input set. Though not designed for it, we can implement updates on the $x$-fast trie in $O(w)$ time by directly updating each level of the dictionary accordingly. Our new predecessor structure, which we call the \emph{$x$tra-fast trie}, instead stores the compact trie of the binary representation of the input set (i.e., the trie where paths of nodes with a single child are merged into a single edge). We store a dictionary representing the prefixes (similar to in the $x$-fast trie) using our parallel dictionary structure of Theorem~\ref{thm:parallel_dictionaries}, but now only for the branching nodes in the compact trie. This reduces the space to $O(n)$. To support predecessor queries for an integer $x$, we generate all $w$ prefixes of $x$ and apply a parallel membership query on these in the dictionary. We show how to identify the longest match in parallel which in turn allows us to identify the predecessor. In total this takes worst-case constant time for the predecessor query. To handle updates, we show how to modify the trie efficiently using scattered memory access operations and a constant number of  dictionary updates, leading to the expected amortized constant time bound of Theorem~\ref{thm:constant_time_predecessor}.

We generalize our result for Theorem~\ref{thm:constant_time_predecessor} to arbitrary $\epsilon > 0$ as follows. The main challenge is that  $\ParallelMember{}$ now supports only $w^\epsilon$ member queries in parallel, so we cannot search for all prefixes of $x$ simultaneously. Instead, we adapt ideas from the $y$-fast trie by Willard~\cite{Willard1983} to our $x$tra-fast trie. The $y$-fast trie works as follows. Partition the input set $S$ into $O(n/w)$ sets $S_1,\ldots,S_t$ where each $S_i$ consists of $w$ consecutive values from $S$, i.e., where $\max(S_i) < \min(S_{i+1})$ for each $i$. Build an $x$-fast trie over the set $S' = \{\max(S_i) \mid i = 1,\ldots,t-1\}$ --- which takes $O(n)$ space since $|S'| = O(n/w)$ --- and a balanced binary search tree over each $S_i$. To determine $\Predecessor{}(x)$, do a predecessor query in the $x$-fast trie to determine the set $S_i$ containing the predecessor of $x$ and do a predecessor query in $S_i$, both of which takes $O(\log w)$ time. Insertions are supported by instead inserting $x$ in $S_i$.  If $S_i$ subsequently becomes too large (e.g., larger than $2w)$, split $S_i$ into two and add an additional element to $S'$ in the $x$-fast trie. This takes $O(w)$ time, which is constant when amortized over the $\Omega(w)$ insertions necessary for $S_i$ to grow too large. Deletions are supported similarly. In our data structure we use dynamic fusion trees by P\u{a}tra\c{s}cu and Thorup~\cite{PT2014} for each $S_i$, which solves the predecessor problem on sets of size $w^{O(1)}$ in linear space and constant time per operation. We build an \emph{uncompacted} $x$tra-fast trie over $S'$, i.e. the $x$tra-fast trie that also includes non-branching nodes. To support fast queries and updates for an integer $x$, we use the scattered memory operations to simulate a $w^\epsilon$-way search (as opposed to a binary search) to find the longest common prefix between $x$ and $S'$. This eliminates a factor $1/w^\epsilon$ of the remaining possibilities per round, leading to a running time of $O(\log_{w^\epsilon} w) = O(1/\epsilon)$, i.e., constant for any fixed $\epsilon$.

In our data structures we only need to store a constant number of ultrawords during the computation. This is important since modern vector processor architectures only have a limited number of ultraword registers.

\subsection{Outline}
In Section~\ref{sec:uwram_model} we describe the UWRAM model of computation and some useful procedures. In Sections~\ref{sec:parallel_multiply_shift} and \ref{sec:parallel_dictionaries} we show how to do parallel hash function evaluation and $w^\epsilon$-parallel dictionaries, proving Theorem~\ref{thm:parallel_dictionaries}. Finally, in Section~\ref{sec:xtra_fast_trie} we prove Theorem~\ref{thm:constant_time_predecessor} for $\epsilon = 1$, which we generalize to arbitrary $\epsilon > 0$ in Section~\ref{sec:predecessor_with_smaller_ultrawords}.
    
\section{The Ultra-Wide Word RAM Model}\label{sec:uwram_model}
The \emph{word RAM} model of computation~\cite{Hagerup1998} consists of an unbounded memory of $w$-bit words and a standard instruction set including  arithmetic, boolean, and bitwise operations (denoted `$\&$', `$|$' and `$\sim$' for \textit{and}, \textit{or} and \textit{not}) and shifts (denoted `$\gg$' and `$\ll$') such as those available in standard programming languages (e.g., C).  We make the standard assumption that we can store a pointer into the input in a single word and hence $w \geq \log n$, where $n$  is the size of the input, and for simplicity we assume that $w$ is even. We denote the address of $x$ in memory as $\addr{x}$, and the address of an array is the address of its first index. The time complexity of a word RAM algorithm is the number of instructions and the space is the number of words stored by the algorithm. 

The \emph{ultra-wide word RAM} (UWRAM) model of computation~\cite{FLNS2015} extends he word RAM model with special \emph{ultrawords} of $w^2$ bits (in Section~\ref{sec:predecessor_with_smaller_ultrawords} we consider the case where ultrawords have  $w^{1+\epsilon}$ bits for any fixed $\epsilon > 0$). As in~\cite{FLNS2015}, we distinguish between the \emph{restricted UWRAM} that supports a minimal set of instructions on ultrawords consisting of addition, subtraction, shifts, and bitwise boolean operations, and the \emph{multiplication UWRAM} that additionally supports multiplications. We extend the notation for bitwise operations and shifts to ultrawords. The UWRAM (both restricted and multiplication) also supports contiguous and scattered memory access operations, as described below. The time complexity is the number of instructions (on standard words or ultrawords) and the space complexity is the number of words used by the algorithms, where each ultraword is counted as $w$ words. The UWRAM model captures (and idealizes) modern vector processing architectures~\cite{Reinders2013, SBBE+2017, CRDI2007}. See also Farzan et al.~\cite{FLNS2015} for a detailed discussion of the applicability of the UWRAM model. 

\begin{figure}
    \centering
    \includegraphics[scale=0.9]{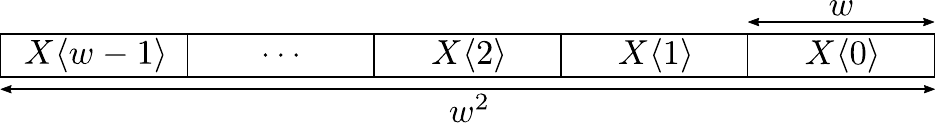}
    \caption{The layout of an ultraword \uw{X}.}
    \label{fig:ultraword_example}
\end{figure}

\subsection{Instructions and Componentwise Operations} 

Recall that ultrawords consists of $w^2$ bits. We often view an ultraword $X$ as divided into $w$ words of $w$ consecutive bits each, which we call the \emph{components} of $X$. We number the components in $X$ from right-to-left starting from $0$ and use the notation $X\angle{i}$ to denote the $i$th word in $X$ (see Figure~\ref{fig:ultraword_example}). We will also use the notation $X = \angle{x_{w-1},\ldots,x_0}$, denoting that $\uwi{X}{i} = x_i$.   

We define a number of useful componentwise operations on ultrawords that we will need for our algorithms in the following. Let $X$ and $Y$ be ultrawords. The \emph{componentwise addition} of $X$ and $Y$, denoted $X + Y$, is the ultraword $Z$ such that $Z\angle{i} = X\angle{i} + Y\angle{i} \mod 2^w$. We define \emph{componentwise subtraction}, denoted $X-Y$,  and \emph{componentwise multiplication}, denoted $XY$, similarly. The \emph{componentwise comparison} of $X$ and $Y$ is the ultraword $Z$ such that $Z\angle{i} = 1$ if $X\angle{i} < Y\angle{i}$ and $0$ otherwise. Given another ultraword $I$ where each component is either $0$ or $1$, we define the \emph{componentwise blend} of $X$, $Y$, and $I$ to be the ultraword $Z$ such that $Z\angle{i} = X\angle{i}$ if $I\angle{i} = 0$ and $\uwi{Z}{i} = Y\angle{i}$ if $I\angle{i} = 1$.

Except for componentwise multiplication, all of the above componentwise operations can be implemented in constant time on the restricted UWRAM using standard word-level parallelism techniques~\cite{Hagerup1998, BGS2022a} (see Appendix~\ref{sec:2wbitmult} for details on blend). For our purposes, we will need componentwise multiplication as an instruction (for evaluating hash functions in parallel) and thus we include this in the instruction set of the UWRAM. This is the UWRAM model that we will use throughout the rest of the paper. Note that all of the componentwise operations are widely supported directly in modern vector processing architectures. For instance, a componentwise multiplication (e.g., the \texttt{vpmullw} operation) is defined in Intel's AVX2 vector extension~\cite{Intel2011}.

We will need componentwise operations on components that are small constant multiples of $w$. In particular, we will need a \emph{$2w$-bit componentwise multiplication} that multiplies $w/2$ components of $w$ bits and returns the $w/2$ resulting components of $2w$ bits. Specifically, let $\uw{X} = \langle 0 , x_{w - 2}, \ldots,0, x_2, 0, x_0 \rangle$ and $\uw{Y} = \langle 0 , y_{w - 2}, \ldots, 0, y_2, 0, y_0 \rangle$, i.e., $X$ and $Y$ store $w/2$ components aligned at the even positions. The $2w$-bit componentwise multiplication is the ultraword $Z = \langle z_{w-2}^+,z_{w-2}^-, \ldots,z_2^+,z_2^-,z_0^+,z_0^-\rangle$ where $z_i^+$ and $z_i^-$ is the leftmost and rightmost $w$ bits, respectively, of the $2w$-bit product of $x_i$ and $y_i$. We can implement $2w$-bit componentwise multiplication using standard techniques in constant time on the UWRAM. See Appendix~\ref{sec:2wbitmult} for details.

Finally, the UWRAM model supports the $\Compress$ operation that, given \uw{X}, returns the word that results from concatenating the rightmost bit of each component of \uw{X}. We do not need the corresponding inverse $\Spread$ operation, defined by Farzan et al.~\cite{FLNS2015}.

\subsection{Memory Access}\label{sec:memory_access} The UWRAM supports standard memory access operations that read or write a single word or a sequence of $w$ contiguous words. More interestingly, the UWRAM also supports \emph{scattered} access operations that access $w$ memory locations (not necessarily contiguous) in parallel. Given an ultraword $A$ containing $w$ memory addresses, a \emph{scattered read} loads the contents of the addresses into an ultraword $X$, such that $X\angle{i}$ contains the contents of memory location $A\angle{i}$. Given ultrawords $X$ and $A$ a  \emph{scattered write} sets the contents of memory location $A\angle{i}$ to be $X\angle{i}$. Scattered memory accesses captures the memory model used in IBM's \emph{Cell} architecture~\cite{CRDI2007}. They also appear (e.g.,  \texttt{vpgatherdd}) in Intel's AVX2 vector extension~\cite{Intel2011}. Scattered memory access operations were also proposed by Larsen and Pagh~\cite{LP2012} in the context of the I/O model of computation. 
Note that while the addresses for scattered writes must be distinct, we can read simultaneously from the same address. We can use this to efficiently copy $x$ into all $w$ components of an ultraword $X$. 
%To do so, we write $x$ to a fixed address in memory and then do a scattered read on an ultraword containing $w$ copies of that address. 
%To do so, write $x$ to memory address $0$ and do a scattered read on $Y \ll w^2 = \angle{0,\ldots,0}$ for any ultraword $Y$. We say that we \emph{load} $x$ into $X$. 
To do so, create the ultraword $\angle{0,\ldots,0}$ by left-shifting any ultraword by $w^2$ bits, write $x$ to address $0$, and do a scattered read on $\angle{0,\ldots,0}$. We say that we \emph{load} $x$ into $X$. 

\section{Computing Multiply-Shift in Parallel}{\label{sec:parallel_multiply_shift}}

We show how to efficiently compute a universal hash function in parallel. The \emph{multiply-shift} hashing scheme is a standard and practically efficient family of universal hash functions due to Dietzfelbinger et al.~\cite{DHKP1997}. For some integer $1 \leq c \leq w$, define the class $H_c = \{h_a  \mid 0 < a < 2^w \text{ and $a$ is odd}\}$ of hash functions where ${h_a(x) = (ax \mod 2^w) \gg (w - c)}$. Each function in $H_c$ maps from $w$-bit to $c$-bit integers. The class $H_c$ is \emph{universal} in the sense that for any $x \neq y$ and for $h_a \in H_c$ selected uniformly at random, it holds that $P[h_a(x) = h_a(y)] \leq 2/2^c$. 

We will show how to evaluate $w$ such functions in constant time. Given $\uwi{X}{i} = x_i$, $\uwi{A}{i} = a_i$ and $\uwi{C}{i} = 2^{c_i}$ where $h_i(x) = (a_ix\mod2^w)\gg(w - c_i)$ the goal is to compute ${\uwi{H}{i}=h_i(x_i)}$. To do so we first evaluate the functions in two rounds of $w/2$ functions each, and then combine the results.

\paragraph*{Step 1: Evaluate the hash function on the even indices.} We construct an ultraword $H_\text{even}$ containing all the values of $h_i(x_i)$ at all even indices $i$. First construct the ultrawords 
\begin{align*}
    C' &= \langle0,2^{c_{w-2}},\ldots,0,2^{c_0}\rangle \\
    T' &= \langle 0, a_{w-2}x_{w-2} \mod 2^w,\ldots,0,a_0x_0\mod 2^w\rangle.
\end{align*}

To do so, we do componentwise multiplication of $C$ with the constant  $M = \angle{0,1,\ldots,0,1}$ and componentwise multiplications of $A$, $X$, and $M$. Then, we do a $2w$-bit multiplication of $C'$ and $T'$ and right shift the result by $w$. This produces the ultraword
\begin{equation*}
    H_{\text{even}} = \angle{\star, (a_{w-2}x_{w-2} \mod 2^w) \gg (w - c_{w-2}),  \ldots, \star,  (a_0x_0\mod 2^w) \gg (w - c_0)}  
\end{equation*}
Thus, all even indices in $H_{\text{even}}$ store the resulting hash values of the integers at the even indices in the input. We will not need the values in the odd indices (resulting from the $2w$-bit multiplication and the right shift) and therefore these are marked with a wildcard symbol $\star$. 

\paragraph*{Step 2: Evaluate the hash function on the odd indices.} Symmetrically, we now construct the ultraword $H_{\text{odd}}$ containing $h_i(x_i)$ at all odd indices $i$. To do so, repeat step~1 and modify the shifting to align the computation for the odd indices. More precisely, right shift \uw{X}, \uw{C} and \uw{A} by~$w$ and repeat step~1, then left shift the result by~$w$ to align the results back to the odd positions. This produces the ultraword
\begin{equation*}
    H_{\text{odd}} = \angle{(a_{w-1}x_{w-1} \mod 2^w) \ll c_{w-1}, \star, \ldots,  (a_1x_1\mod 2^w) \ll c_1, \star}  
\end{equation*}

\paragraph*{Step 3: Combine the results.} Finally, we combine the results by blending $H_\text{even}$ and $H_\text{odd}$ using $\uw{I} = \angle{1,\ldots,1} - \uw{M}$, producing the ultraword $H$ of the even indices of $H_\text{even}$ and the odd indices of $H_\text{odd}$. 
\medskip

This takes constant time since componentwise multiplication, $2w$-bit multiplication, shifting, blending, loading $1$ into $\angle{1,\ldots,1}$, and componentwise subtraction all run in constant time. Hence, we can evaluate each case of $w/2$ hash functions in constant time and combine the results in constant time. In summary, we have the following result. 
\begin{lemma}\label{lemma:parallel_multiply_shift}
Given $\uwi{X}{i} = x_i$, $\,\uwi{A}{i} = a_i$, $\,\uwi{C}{i} = 2^{c_i}$, and the constant $M = \angle{0,1,\ldots,0,1}$ we can evaluate each of the $w$ multiply-shift hash functions $h_i(x) = (a_ix \mod 2^w) \gg (w - c_i)$ by computing the ultraword $\uw{H} = \angle{h_{w-1}(x_{w-1}),\ldots,h_0(x_0)}$ in constant time on a UWRAM.  
\end{lemma}

\section{The \texorpdfstring{$\boldsymbol{w^\epsilon}$}{w\^{}epsilon}-Parallel Dictionary}\label{sec:parallel_dictionaries}
We now show how to construct the $w^\epsilon$-parallel dictionary of Theorem~\ref{thm:parallel_dictionaries}. Throughout the section we assume that $\epsilon = 1$, but the result generalizes to any $\epsilon > 0$ in a straight forward manner. Our data structure is based on a dictionary by Dietzfelbinger et al. that implements a dynamic perfect hashing strategy~\cite{DKMHRT1994}. Their dictionary already supports \Insert{} and \Delete{} in amortized expected constant time. Furthermore, it supports sequential \Member{} queries (i.e. ``is $x \in S$'') in worst case constant time. We will show that we can use scattered memory operations to run $w$ \Member{} queries simultaneously, thus implementing \ParallelMember{} in constant time.

\subsection{Dynamic Perfect Hashing}\label{sec:parallel_dictionaries_data_structure}

In this section we briefly describe the contents of the data structure of Dietzfelbinger et al.~\cite{DKMHRT1994}. Note that we use the multiply-shift hashing scheme, while they use another class of universal hash functions. Multiply-shift satisfies all the necessary constraints and the analysis from~\cite{DKMHRT1994} still works. It does however incur a multiplicative, constant space overhead for our arrays since the range of a multiply-shift function is a power of two.

The main idea of the data structure is as follows. Let $S$ be a set of $w$-bit integers. Choose $h \in H_c$ and partition $S$ into $2^c = \Theta(n)$ sets $S_0,\ldots,S_{2^c - 1}$ where $S_i = \{x \mid x \in S \text{ and } h(x) = i\}$. Each set 
$S_i$ is stored in a separate array using a hash function $h_i$. Dietzfelbinger et al.\ show how to implement the operations \Insert{} and \Delete{} such that they
maintain that $h_i$ has no collisions on $S_i$.

The data structure consists of the following.
\begin{itemize}
    \item For each $S_i$, store an array $T_i$ of size $2^{c_i}$. Let $h_i(x) = (a_ix\mod 2^w) \gg (w - c_i)$. For each $x \in S_i$ let $T_i[h_i(x)] = x$, i.e.  the position that $x$ hashes to stores $x$. If there is no $x \in S_i$ that hashes to $j$, then $T_i[j] = 2^{w-1}$ if $j = 0$ and $T_i[j] = 0$ otherwise. We claim that $h_i(0)$ is always zero and $h_i(2^{w-1})$ is never zero, so it follows from this construction that $x \in S_i$ if and only if $T_i[h_i(x)] = x$. We have that $h_i(2^{w-1})$ is not zero because
    \[
        h_i(2^{w-1}) \;\;=\;\; (a_i2^{w-1} \mod 2^w) \gg (w - c_i) \;\;=\;\; 2^{w-1} \gg (w - c_i) \;\;\geq\;\; 1.
    \]
    The second step follows since $a_i$ is odd; then $a_i2^{w-1} = 2^{w-1} + (a_i - 1)2^{w-1}$, and the latter term is $0$ modulo $2^w$ since $a_i - 1$ is even. The last step follows because $c_i \geq 1$. 

    \item An array $T$ of size $2^c$. At index $T[i]$ we store the $5$-tuple $(\addr{T_i}, 2^{c_i}, a_i, \star, \star)$ where $\star$ are book-keeping values used by \Insert{} and \Delete{}. Note that $2^{c_i}$ and $a_i$ encode $h_i$. 
    
    \item The integers $a$ and $2^c$ representing the top-level hash function $h(x) = (ax \mod 2^w) \gg (w - c)$, as well as \addr{T}. 
\end{itemize}

It follows from this construction that $x \in S$ if and only if $T_i[h_i(x)] = x$ where $i = h(x)$. Dietzfelbinger~et~al. show that the data structure uses linear space, that \Member{} runs in worst-case constant time, and that \Insert{} and \Delete{} run in amortized expected constant time~\cite{DKMHRT1994}. 

\paragraph*{Extending the Data Structure.}
We extend this data structure by storing the constant $\uw{M} = \langle 0,1,\ldots,0,1,0,1\rangle$ from Section~\ref{sec:parallel_multiply_shift} used to evaluate multiply-shift functions in parallel. This increases the space of the data structure to $O(n+w)$. Note that linear space in $w$ is needed even to store the input to a $\ParallelMember$ query.

\subsection{Parallel Queries}
\label{sec:parallel_dictionaries_parallel_queries}
In this section, we begin by describing a single $\Member$ query, before we show how to run~$w$ copies of the $\Member$ query in parallel to support $\ParallelMember$. We compute $\Member{}(x)$ as follows.
\begin{enumerate}
    \item Using $a$ and $2^c$, compute $j = h(x)$.
    \item Let $q = \addr{T} + 5j = \addr{T[j]}$ (recall that each index in $T$ stores five words). Read the values stored at $q$, $q + 1$ and $q + 2$ to get respectively $\addr{T_j}$, $2^{c_j}$ and $a_j$, the first three words stored at $T[j]$. Compute $k = h_j(x)$.
    \item Check whether the value stored at  $\addr{T_j} + k = \addr{T_j[k]}$ is equal to $x$.
\end{enumerate}
The parallel algorithm runs this algorithm for all $w$ inputs simultaneously. Given $\uw{X} = \angle{x_{w-1}, \ldots, x_0}$ we implement $\ParallelMember{}(\uw{X})$ as follows. Each of the steps below executes the corresponding step above in parallel for each of the~$w$ inputs. 

\paragraph*{Step 1: Evaluate the top-level hash function.}  Load the two ultrawords $\uw{A} = \angle{a,\ldots,a}$ and $\uw{C} = \angle{2^c, \ldots, 2^c}$.  Compute the ultraword 
%\[
    $J = \angle{h(x_{w-1}), \ldots, h(x_0)}$
%\]
using the multiply-shift algorithm of Lemma~\ref{lemma:parallel_multiply_shift}. 

\paragraph*{Step 2: Evaluate each of the second-level hash functions.}Load $F = \angle{5, \ldots, 5}$ and $\uw{P} = \angle{\addr{T}, \ldots, \addr{T}}$. Compute $\uw{Q} = P + FJ$. Then $\uwi{Q}{i} = \addr{T} + 5\uwi{J}{i} = \addr{T[\uwi{J}{i}]}$. Do scattered reads of $Q$, $Q + \angle{1, \ldots, 1}$, and $Q + \angle{2, \ldots, 2}$ to produce the ultrawords $P'$, $C'$, and $A'$. We have that
\begin{align*}
    P' &= \langle \addr{T_{\uwi{J}{w-1}}},\ldots,\addr{T_{\uwi{J}{0}}}\rangle \\
    C' &= \langle 2^{c_{\uwi{J}{w-1}}},\ldots,2^{c_{\uwi{J}{0}}}\rangle\\
    A' &= \langle a_{\uwi{J}{w-1}},\ldots,a_{\uwi{J}{0}}\rangle
\end{align*}
Compute the ultraword
%\begin{align*}
    $K = \angle{h_{\uwi{J}{w-1}}(x_{w-1}), \ldots, h_{\uwi{J}{0}}(x_0)} $
%\end{align*}
using the multiply-shift algorithm of Lemma~\ref{lemma:parallel_multiply_shift}.

\paragraph*{Step 3: Check whether the inputs are present in the dictionary.}
Do a scattered read of \uw{P' + K} and name the result \uw{R}. Then $\uwi{R}{i} = T_{j}[h_j(x_i)]$ where $j = h(x_i)$.  Return the result \uw{I} of componentwise equality between \uw{X} and \uw{R}. That is
\[
\uwi{I}{i} = \begin{cases}
    1 & \text{ if } \uwi{X}{i} = \uwi{R}{i}\\
    0 & \text{ otherwise}
\end{cases}
\]

Evaluating the hash functions in steps~1 and 2 takes constant time according to Lemma~\ref{lemma:parallel_multiply_shift}. The remaining operations are scattered reads, loads and componentwise operations, all of which run in constant time. Since there is only a constant number of operations, \ParallelMember{} runs in constant time. This concludes the proof of Theorem~\ref{thm:parallel_dictionaries}. 

Note that both the algorithm for parallel hashing and the dictionary generalizes to the case with $w^{1+\epsilon}$-bit ultrawords and $w^\epsilon$ inputs in a straight forward manner. In this case, the space is $O(n + w^\epsilon)$ since the ultraword constants use only $w^\epsilon$ space.

\subsection{Satellite Data}
Suppose we associate some value $\data{x}$ with each $x \in S$. We extend the data structure to support the following operation, where $\uw{X} = \angle{x_{w-1},\ldots,x_0}$ as above.
\begin{itemize}
    \item $\Retrieve{}(\uw{X})$: returns a pair $(\uw{I}, \uw{D})$ where \uw{I} is the result of $\ParallelMember(\uw{X})$ and 
\[
    \uwi{D}{i} = \begin{cases}
        \addr{\data{x_i}} & \text{ if } \uwi{I}{i} = 1\text{, i.e if } x_i \in S\\
        \text{undefined} & \text{ otherwise}
    \end{cases}
\] 
\end{itemize} 
We return $\addr{\data{x}}$ instead of $\data{x}$ since the data would not fit into an ultraword if $\data{x}$ requires more than one word to store. 

We extend the data structure as follows to support \Retrieve{}. Store two words for each index in $T_i$. For each $x \in S_i$, the first word in $T_i[h_i(x)]$ stores $x$ and the second stores $\addr{\data{x}}$. The remaining entries store either $0$ or $2^{w-1}$, as above.

To do the retrieval, first compute $\uw{I} = \ParallelMember{}(\uw{X})$. However, in step~3, multiply $K$ by $\angle{2,\ldots,2}$ before the scattered read since each index in $T_i$ now stores two words. Also, add $\angle{1,\ldots,1}$ to  \uw{P' + \angle{2,\ldots2}K} and do a scattered read to compute the ultraword $\uw{D}$. The space of the data structure remains $O(n + w)$ (assuming that $\data{x}$ uses constant space), and \Retrieve{} runs in constant time.

\section{The \texorpdfstring{$\boldsymbol{x}$}{x}tra-fast Trie}\label{sec:xtra_fast_trie}

In this section we prove Theorem~\ref{thm:constant_time_predecessor} for the special case where $\epsilon = 1$, i.e. where ultrawords consist of $w^2$ bits. We generalize our result to arbitrary $\epsilon > 0$ in Section~\ref{sec:predecessor_with_smaller_ultrawords}. Our data structure, the $x$tra-fast trie, supports \Predecessor{} in worst case constant time and \Insert{} and \Delete{} in amortized expected constant time. In our description we assume that we have keys of $w-1$ bits each and we give a solution that uses $O(n + w)$ space. At the end of this section we will reduce the space to $O(n)$ and extend the solution to $w$-bit keys, proving Theorem~\ref{thm:constant_time_predecessor} for $\epsilon = 1$.

\subsection{Data Structure}

Consider the compacted trie $T$ over the binary representation of the elements in $S$. For each node $v \in T$ define $\str(v)$ to be the bitstring encoded by the path from the root to $v$ in $T$. Also let $\minl(v)$ and $\maxl(v)$ be the smallest and largest leaves in the subtree of $v$, respectively. By $\minl(v)$ and $\maxl(v)$ we refer both to a leaf and to the value the leaf represents. 

\begin{figure}
\centering
\includegraphics[scale=1.4]{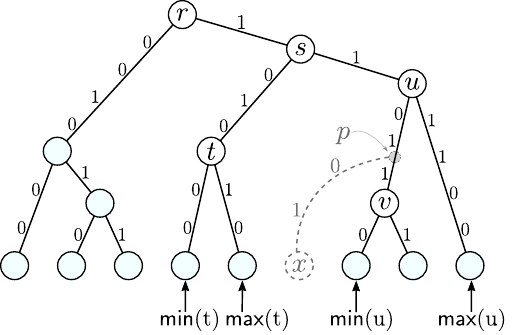}
    \caption{An $x$tra-fast trie for $S = \{001000, 001010, 001011, 101000,101010,110110, 110111, 111100\}$. The dashed edge and nodes illustrate how the trie would change if $x = 110101$ were inserted. The exit edge for $x$ is $(u,v)$ since we match the bitstring $1101$ but do not match the next $1$ on $(u,v)$. Similarly, the exit edge for $100100$ is $(s,t)$. We have that $\key(u,v) = \lab(u,v)\underline{1000} = 110\underline{1000}$ where the underlined part is what we append to the labels to disambiguate the keys. Similarly, $\key(r,s) = 1\underline{100000}$ and $\key(s,t) = 10\underline{10000}$. The dictionary entry of $(s,u)$ has $\key(s,u) = 11\underline{10000}$, and the min- and max-pointer of $(s,u)$ are $\addr{\minl(u)}$ and $\addr{\maxl(u)}$. Similarly, the min-pointer of $(r,s)$ is to $\minl(s) = \minl(t)$ and the max-pointer is to $\maxl(s) = \maxl(u)$. Note that if we insert $x$ we would have to update the min-pointer of $(s,u)$, since $x < \minl(v)$. However, the min-pointer of $(r,s)$ remains unchanged since $\minl(t) < x$.} 
    \label{fig:xtra_fast_trie_example}
\end{figure}

For each edge $(u,v) \in T$, let $\lab(u,v)$ be $\str(u)$ followed by the first bit on the edge $(u,v)$. Define $\key(u,v)$ to be $\lab(u,v)$ followed by a single $1$-bit and $w - |\lab(u,v)| - 1$ zeroes. Note that $|\key(u,v)| = w$ and that the keys of two distinct edges in $T$ always differ. See Figure~\ref{fig:xtra_fast_trie_example} for an example.

We define the \emph{exit edge} for an integer $x$ to be the edge in $T$ where the match of $x$ ends. In other words, it is the edge $(u,v) \in T$ such that $\lab(u,v)$ is a prefix of $x$ and $|\lab(u,v)|$ is maximum. See Figure~\ref{fig:xtra_fast_trie_example} for an example. It is possible that $x$ has no exit edge if the root has fewer than two children. 

Our data structure consists of the following: 
\begin{itemize}
    \item A sorted, doubly linked list $L$ of the leaves of $T$, i.e., the elements of $S$. 
    \item A dictionary $D$ supporting parallel queries using Theorem~\ref{thm:parallel_dictionaries}. 
    For each edge $(u,v) \in T$ we store an entry in $D$ with the key $\key(u,v)$ and $\data{u,v} = (\addr{\minl(v)}, \addr{\maxl(v)})$. Here, $\addr{\minl(v)}$ and $\addr{\maxl(v)}$ are the addresses to the corresponding elements in $L$, and we denote the addresses to $\minl(v)$ and $\maxl(v)$ as the \emph{min-} and \emph{max-pointer} of $(u,v)$.
    \item The two ultraword constants \uw{M'} and \uw{H} described in the next section.
\end{itemize}
Storing $L$ and the ultraword constants takes $O(n+w)$ space combined. Since $T$ is compacted there are $O(n)$ entries in $D$, so by Theorem~\ref{thm:parallel_dictionaries} the dictionary also uses $O(n + w)$ space.    

\subsection{Predecessor Queries}{\label{sec:predecessor_query}}
The main idea of the predecessor query for $x$ is to first find the exit edge of $x$ by simultaneously searching for all prefixes of $x$ in $D$. Then we use the min- and max-pointer of the exit edge to find the predecessor of $x$. If $x$ has no exit edge, then the root does not have an outgoing edge matching the leftmost bit of $x$. If the leftmost bit of $x$ is $1$, the predecessor of $x$ is the largest leaf in the left subtree of the root, and otherwise $x$ has no predecessor. Assuming that $x$ has an exit edge, the procedure has three steps.\medskip

\paragraph*{Step 1: Compute all prefixes of $x$.}  Let $b_{w-2}b_{w-3}\cdots b_{0}$ be the binary representation of $x$ of length $w-1$. We  compute the ultraword
\[\uw{\overline{X}} = \angle{b_{w-2}b_{w-3}\cdots b_{0}1\;\,,\;\, b_{w-2} b_{w-3}\cdots b_1 1 0 \;\,,\;\, \ldots\;\,,\;\,10\cdots 0}.\]
That is,
$\uwi{\overline{X}}{i}$ contains the prefix of $x$ of length $i$ followed by a $1$-bit and $w - i - 1$ zeroes. Thus, for any edge $(u,v) \in T$ such that $\lab(u,v)$ is the length-$i$ prefix of $x$, we have $\uwi{\overline{X}}{i} = \key(u,v)$. We compute~$\uw{\overline{X}}$ as follows.

Let \uw{M'} be the constant such that \uwi{M'}{i} consists of $i$ consecutive $1$-bits followed by $w-i$ consecutive $0$-bits. 
Let \uw{H} be the constant where the $(i+1)$th leftmost bit in \uwi{H}{i} is $1$ and the remaining bits are zeroes.
First load $x$ into \uw{X} such that $\uw{X} = \langle x,x,\ldots,x\rangle$. Then compute 
$\uw{\overline{X}} = (\uw{X}~\&~\uw{M'}) \mid \uw{H}$.

\paragraph*{Step 2: Find the exit edge $(u,v)$ of $x$.} First do $(\uw{I}, \uw{P}) = \Retrieve{}(\,\uw{\overline{X}}\,)$ on $D$. 
Then compute $c =\Compress(\uw{I})$ such that the $i$th rightmost bit in $c$ is $1$ if $\uwi{I}{i} = 1$ and zero otherwise. Note that $x$ has no exit edge if $c = 0$. Find the index $k$ of the leftmost bit in $c$ that is $1$ (see~\cite{FW1993}). Then $\uwi{\overline{X}}{k} = \key(u,v)$ where $(u,v)$ is the exit edge of $x$. Furthermore, the values stored at the addresses \uwi{P}{k} and $\uwi{P}{k} + 1$ are the min- and max-pointers of $(u,v)$, respectively.

\paragraph*{Step 3: Find the predecessor of $x$.} Use the min- and max-pointer of $(u,v)$ found in step~2 to retrieve $\minl(v)$ and $\maxl(v)$. If $x \geq \maxl(v)$ then return $\maxl(v)$, otherwise return the element immediately left of $\minl(v)$ in $L$. Note that there might not be an element immediately left of $\minl(v)$ if $x$ is smaller than than everything in $S$, in which case $x$ has no predecessor.\medskip

Since we search for all prefixes of $x$ and take the edge corresponding to the longest prefix found, we 
find the exit edge $(u,v)$ of $x$. If $x \in S$, then $x = v = \maxl(v)$ and we correctly return that $x$ is the predecessor of itself. If $x \not\in S$ then the path to where $x$ would have been located if it were in $T$ branches off $(u,v)$ either to the left (if $x < \minl(v)$) or right (if $x > \maxl(v)$). In the first case, $\Predecessor{}(x)$ is the element located immediately left of $\minl(v)$ in $T$, and in the second case $\Predecessor{}(x)$ is $\maxl(v)$.

By Theorem~\ref{thm:parallel_dictionaries} the parallel dictionary query in step 2 takes worst case constant time. Finding the leftmost bit that is $1$ takes constant time on the word RAM~\cite{FW1993}. The remaining operations are standard operations available in the model, so the procedure runs in constant time. 

\subsection{Insertions} \label{sec:insertions}
The main idea of the insertion procedure is as follows. Since $T$ is compacted, inserting a new leaf $x$ will cause only a constant number of edges to be inserted and removed, so we can make these changes sequentially. Furthermore, some of the at most $w-1$ edges on the path from the root to $x$ might have their min- or max-pointers changed, and we will update these edges in parallel. 

Consider inserting $x = 110101$ in the trie in Figure~\ref{fig:xtra_fast_trie_example}. When $x$ is inserted we add a new leaf for $x$, as well as a new node $p$ at the location where the path to $x$ branches off the exit edge $(u,v)$ of $x$. This removes the edge $(u,v)$, but adds the three new edges $(u,p)$, $(p,x)$ and $(p,v)$. Furthermore, we must update the min-pointer of $(s,u)$, because $\minl(v)$ was replaced by $x$ as the smallest leaf under $u$. On the other hand, we do not update the min-pointer of $(r,s)$ because $\minl(t)$ is smaller than $x$.  Note that we do not explicitly store internal nodes and therefore do not add $p$ anywhere in the data structure. 

We now describe the insertion procedure. First we note that if $x$ does not have an exit edge it is because the root does not have an outgoing edge which shares the same leftmost bit as $x$. This case is easily solved by adding an edge from the root to the new leaf $x$ and adding $x$ to either the start or end of $L$. We will now assume that $x$ has an exit edge, and also that $x$ branches off its exit edge to the left; the other case is symmetric.  

\paragraph*{Step 1: Find the predecessor of $x$.} Do a predecessor query as described in Section~\ref{sec:predecessor_query}, which determines% The predecessor of $x$ in $L$,  the exit edge $(u,v)$ of $x$, along with $\lab(u,v)$ and $\data{u,v} = (\addr{\minl(v)}, \addr{\maxl(v)})$, and the result $(\uw{I},\uw{P})$ of $\Retrieve{}(\,\uw{\overline{X}}\,)$ on $D$.
\begin{itemize}
    \item The predecessor of $x$ in $L$.
    \item The exit edge $(u,v)$ of $x$,  $\lab(u,v)$ and $\data{u,v} = (\addr{\minl(v)}, \addr{\maxl(v)})$.
    \item The result $(\uw{I},\uw{P})$ of $\Retrieve{}(\,\uw{\overline{X}}\,)$ on $D$.
\end{itemize}

\paragraph*{Step 2: Insert $x$ in $L$.} Insert $x$ immediately to the right of its predecessor in $L$. 

\paragraph*{Step 3: Update edges.}\label{p:a} We insert $(u,p)$, $(p,x)$ and $(p,v)$ and remove $(u,v)$ from $D$. We find the labels of the three edges to insert as follows. We have that $\lab(u,p) = \lab(u,v)$ since $(u,p)$ is  the edge $(u,v)$ shortened by adding the node $p$ and since only the first character of the edge affects the label. By definition, $\lab(p,x)$ and $\lab(p,v)$ consist of $\str(p)$ with a zero and a one appended, respectively. We compute $\str(p)$ by finding the longest common prefix $\hat{p}$ of $x$ and $\minl(v)$. To do so, do bitwise XOR between $x$ and $\minl(v)$ and find the index $k$ of the leftmost bit that is $1$ in the result (see~\cite{FW1993}). Now $k$ indicates the leftmost bit where $x$ and $\minl(v)$ differ. To extract the longest common prefix compute  $\hat{p} = x~\&~{\sim}((1\ll(k+1))-1)$. 
Given the labels we can easily construct the keys for the edges. 

We now construct the satellite data for the edges. Both the min- and max-pointer for $(p,x)$ are $\addr{x}$ since $x$ is a leaf. For $(p,v)$ they are $\addr{\minl(v)}$ and $\addr{\maxl(v)}$, which were determined during the predecessor query. Finally, the min-pointer for $(u,p)$ is $\addr{x}$ and the max-pointer is $\addr{\maxl(v)}$.

\paragraph*{Step 4: Update min-pointers.} We update the min-pointers for the edges on the path from the root to $u$ that are incorrect after inserting $x$. Note that inserting $x$ cannot invalidate any max-pointers since we assumed that $x$ branched off its exit edge to the left. The edges that must be updated are exactly those that have a min-pointer to $\minl(v)$, since $x$ has replaced $\minl(v)$ as the smallest leaf under $u$. 

Consider the result $(\uw{I}, \uw{P})$ from the \Retrieve{} query. We begin by setting $\uwi{I}{k'} = 0$ for the index~$k'$ corresponding to the exit edge $(u,v)$ of $x$ (we know $k'$ from the predecessor query). The indices in \uw{I} that now contain $1$ indicate the edges on the path from the root to $u$.
 
Next we identify the edges that needs to be updated by creating \uw{I'} where $\uwi{I'}{i} = 1$ if and only if both $\uwi{I}{i} = 1$ and what is stored at address \uwi{P}{i} is the address of $\minl(v)$. To do so, first do a scattered read of \uw{P} and store the result in \uw{M}. Now $M$ contains $\addr{\minl(b)}$ for each edge $(a,b)$ on the path to $u$.\footnote{If $x$ branched off to the right of its exit edge, we would do a scattered read of $\uw{P} + \angle{1,\ldots,1}$ to load the max-pointers instead of min-pointers.} Note the value  of \uwi{P}{i} is arbitrary if $\uwi{I}{i} = 0$, i.e. if no edge has the length-$i$ prefix of $x$ as its label. Load $\addr{\minl(v)}$ into the ultraword \uw{V}. Let \uw{E} be the result of componentwise equality between \uw{M} and \uw{V}. Then $\uwi{E}{i} = 1$ if and only if what is stored at address \uwi{P}{i} is $\addr{\minl(v)}$. Finally compute $\uw{I'} = \uw{I}\;\&\;\uw{E}$.

Now we use \uw{P} and \uw{I'} to update the incorrect min-pointers. First, load the address of the node for $x$ into \uw{U}. Then compute \uw{B} by blending \uw{M} (the result of the scattered read of \uw{P}) and \uw{U} conditioned on \uw{I'} such that
\[
    \uwi{B}{i} = 
        \begin{cases}
            \uwi{M}{i} & \text{ if } \uwi{I'}{i} = 0 \quad \text{(i.e. the value already at the address \uwi{P}{i}) }\\
            \uwi{U}{i} & \text{ if } \uwi{I'}{i} = 1  \quad \text{(i.e. the address of $x$) }\\
        \end{cases}
\] 

Finally, do a scattered write of \uw{B} to the addresses in \uw{P}. Hence, what is stored at the address \uwi{P}{i} remains the same if $\uwi{I'}{i} = 0$ and is replaced by the address of $x$ otherwise. \medskip

The predecessor query in step~1 takes constant time. The operations in step~2 and step~4 are all standard RAM or UWRAM operations, except for finding the leftmost $1$-bit which takes constant time~\cite{FW1993}. The dictionary updates in step~3 run in amortized expected constant time by Theorem~\ref{thm:parallel_dictionaries}. Since the rest of step~3 consists of standard operations, the running time for insertions is amortized expected constant. 

\subsection{Deletions}
The deletion procedure is essentially the inverse of the insertion procedure. We assume that $x$ is the left child of its parent $p$; the other case is symmetric.

\paragraph*{Step 1: Find $x$.} Do a predecessor query for $x$. Since $x \in S$, the predecessor of $x$ is itself. This determines
\begin{itemize}
    \item The position of $x$ in $L$. 
    \item The exit edge $(p,x)$ for $x$, along with $\lab(p,x)$. Since $x \in S$, this edge must end in the leaf for $x$. 
    \item The result $(\uw{I},\uw{P})$ of $\Retrieve{}(\,\uw{\overline{X}}\,)$ on $D$.
\end{itemize}

\paragraph*{Step 2: Update min-pointers.} If $p$ is the root (i.e. if $|\lab(p,x)| = 1$) we remove the edge $(p,x)$ from $D$ and remove $x$ from $L$ which completes the deletion of $x$. Otherwise $p$ is an internal node and must have another child which we denote by $v$. Consider the edges on the path to $p$. Any min-pointer to $x$ should be replaced by the address of $\minl(v)$, since $\minl(v)$ is the successor of $x$ and also in the subtree of all of these edges. We find $\minl(v)$ in the node immediately right of $x$ in $L$. As we did for insertions, replace any min-pointer that is an address of $x$ by the address of $\minl(v)$ in parallel using \uw{I} and \uw{P}.

\paragraph*{Step 3: Delete edges.} We delete $(p,x)$ and $(p,v)$ from $D$. Determine $\lab(p,v)$ by flipping the last bit in $\lab(p,x)$. Using the labels we easily find the keys. Note that we do not explicitly delete the edge $(u,p)$ or insert the edge $(u,v)$. These two edges share the same key, and the min-pointer of $(u,p)$ was changed to the address of $\minl(v)$ in step~2.

\paragraph*{Step 4: Update $\mathbf{L}$.} Remove $x$ from $L$.\medskip

Steps~1, 2 and 4 all take constant time (see Sections~\ref{sec:predecessor_query} and \ref{sec:insertions}). The two deletions in step~3 take amortized constant time according to Theorem~\ref{thm:parallel_dictionaries}. The remainder of step~3 takes constant time, so deletions run in amortized expected constant time.

\subsection{Reducing to Linear Space and Supporting \texorpdfstring{$\boldsymbol{w}$}{w}-bit Keys}{\label{sec:linear_space_and_w_bit_keys}}
Here, we reduce the space to $O(n)$ and show how to support $w$-bit keys, concluding the proof of Theorem~\ref{thm:constant_time_predecessor}.

The $O(w)$ term in the space bound above is due to the $w^\epsilon$-parallel dictionary $D$ and $O(1)$ ultraword constants. To avoid this when $n = o(w)$, we will initially support $\Predecessor$, $\Insert$ and $\Delete$ using the \textit{dynamic fusion tree} by P\u{a}tra\c{s}cu and Thorup~\cite{PT2014} (based on the fusion tree by Fredman and Willard~\cite{FW1993}), which uses linear space and supports all three operations in constant time for sets of size $w^{O(1)}$. Simultaneously, we build the ultraword constants we need over the course of $\Theta(w)$ insertions, maintaining linear space. When $n \geq w$, the constants have been built and we move all elements into the trie. If at any point $n \leq w/2$, we move all elements from the trie into  a fusion tree and remove the trie and the ultraword constants, leaving us with linear space and $\Theta(w)$ \Insert{} operations in which to rebuild the constants. Updates still run in amortized expected constant time since we always do $\Omega(w)$ updates before we move $O(w)$ elements. 

To extend the solution to work with $w$-bit keys, we partition the input set $S$ into $S_0$ and $S_1$ where $S_i = \{s \mid s \in S \text{ and the leftmost bit of $s$ is } i\}$, and store an $x$tra-fast trie for each set. Suppose the leftmost bit of an integer $x$ is $i$. An \Insert{}, \Delete{} or \Predecessor{} operation on $x$ is performed on the data structure for $S_i$. Additionally, if $i = 1$ and the \Predecessor{} query on $S_1$ returns that $x$ has no predecessor, we return the largest element in $S_0$, or report that $x$ has no predecessor if $S_0$ is empty. 

\section{The \texorpdfstring{$\boldsymbol{x}$}{x}tra-fast Trie With Smaller Ultrawords}
\label{sec:predecessor_with_smaller_ultrawords}
In this section we show how to match the bounds of Theorem~\ref{thm:constant_time_predecessor} when ultrawords consist of only $w^{1+\epsilon}$ bits (i.e. $w^\epsilon$ components) for any fixed $\epsilon > 0$. The model is otherwise exactly as described in Section~\ref{sec:uwram_model}. 

As mentioned, our data structure based on the $y$-fast trie by Willard~\cite{Willard1983} (see Section~\ref{sec:techniques}). We partition the input set $S$ into $O(n/w)$ sets $S_1,\ldots,S_t$ where each $S_i$ consists of $w$ consecutive values from $S$, i.e., where $\max(S_i) < \min(S_{i+1})$ for each $i$ (note that $|S_t| < w$ is possible). We build a dynamic fusion tree~\cite{PT2014} over each $S_i$ and an \emph{uncompacted} $x$tra-fast trie $\mathcal{T}$ over $S'$, i.e., the $x$tra-fast trie where we include non-branching nodes. The size of $S'$ is $O(n/w)$ and each root-to-leaf path has length $O(w)$, so storing the uncompacted trie uses $O(n + w^\epsilon)$ space, where the additional $w^\epsilon$ is due to the $w^\epsilon$-parallel dictionary. We also store a collection $B$ of ultraword constants (to be described shortly) that increases the space to $O(n + w)$. Note that we use the same trick as in Section~\ref{sec:linear_space_and_w_bit_keys} to reduce this to linear in $n$.

We answer $\Predecessor{}(x)$ as in the $y$-fast trie by first determining the predecessor of $x$ in $\mathcal{T}$, and then finding the predecessor in the corresponding dynamic fusion tree, the latter of which takes constant time~\cite{PT2014}. We show that we can find the longest common prefix between $x$ and $\mathcal{T}$ in constant time, from which it follows that we can find the predecessor of $x$ in $\mathcal{T}$ in constant time (see Section~\ref{sec:predecessor_query}). In the $y$-fast trie this is done by binary searching over the binary representation of $x$, taking $O(\log w)$ time. We speed up the process by doing a $w^\epsilon$-way search instead, reducing the running time to $O(\log_{w^\epsilon} w) = O(1/\epsilon)$, or constant for any fixed $\epsilon$. To do so, we first construct the ultraword $\uw{X_R}$ that contains the labels corresponding to the prefixes of $x$ of length $w^{1-\epsilon},2w^{1-\epsilon},\ldots,w^\epsilon w^{1-\epsilon}$. We then do a $\ParallelMember{}$ query in the dictionary for $\mathcal{T}$, $\Compress{}$ the resulting ultraword (yielding a word indicating which labels were found), and find the most significant bit to determine the longest prefix found. This eliminates all but $w/w^\epsilon$ prefixes as candidates for the longest common prefix with $\mathcal{T}$, and we recurse on this range. To construct the correct labels we use the ultraword constants in $B$. Recall that in Section~\ref{sec:predecessor_query} we use the constants $M'$ and $H$ to compute the labels for the parallel member query by $\uw{\overline{X}} = (\uw{X}~\&~\uw{M'}) \mid \uw{H}$. We can compute any collection of $w^\epsilon$ prefix-labels of $x$ in this way, provided that we use the correct constants. We let $B$ encode a B-tree of degree $\Theta(w^\epsilon)$ over $M'$ and $H$, allowing us to perform the $w^\epsilon$-way search. Consider some node $v$ in $B$ that has $k+1$ children. In $v$ we store $k$ of the values from $M'$ in an array $M'_v$ and the corresponding $k$ values from $H$ in another array $H_v$, ensuring that $k \leq w^\epsilon$ so that each array fits into an ultraword. We additionally store $k$ and the pointers to the children of $v$. When we visit $v$ during the search, we read $M'_v$ and $H_v$ into the $k$ least significant components of two ultrawords. If $k < w^\epsilon$, the $w^{1+\epsilon} - kw$ most significant bits of these ultrawords will contain some values that are irrelevant to the parallel member query; we zero out these bits by doing bitwise $\&$ with $(1 \ll kw) - 1$. This does not cause false positives to occur in the $\ParallelMember{}$ query since no edges in $\mathcal{T}$ has the label $0$ due to how labels are constructed. We then compute the $k$ prefix-labels of $x$, do the parallel lookup in the dictionary, $\Compress{}$ the result, and find the most significant bit to determine which child of $v$ to continue the search in. Since $B$ is a B-tree over $O(w)$ values it uses $O(w)$ space. Furthermore, the height of the tree is $O(1/\epsilon)$ since the branching factor is $\Theta(w^\epsilon)$. We use constant time per node, concluding the proof of the predecessor query.

We also support insertions as in the $y$-fast trie. We determine which set $S_i$ to add the new element to and update that dynamic fusion tree in constant time. If $S_i$ becomes too large we split it (by deleting and reinserting each element in another dynamic fusion tree) and add a separator element to $\mathcal{T}$. This takes expected $O(w)$ time in total (the expectation is from adding at most $w$ new edges to the $w^\epsilon$-parallel dictionary), which is expected constant when amortized over the $\Omega(w)$ updates between splits. Deletions are supported similarly.

\section{Conclusion and Open Problems}
We have studied the predecessor problem on the UWRAM model of computation. We have given a linear space data structure that supports predecessor queries in worst case constant time and updates in amortized expected constant time, even when ultrawords consist of only $w^{1+\epsilon}$ bits for any fixed $\epsilon > 0$.

Furthermore, we have shown how to implement a $w^\epsilon$-parallel dictionary on the UWRAM. The dictionary supports $w$ (or $w^\epsilon$) simultaneous membership queries in worst case constant time and individual updates in amortized expected constant time. 

We wonder if it is possible to achieve constant time with high probability for all operations in the predecessor problem. The limiting factor for our solution is the time for updates in the $w^\epsilon$-parallel dictionary. There are dictionaries that achieve constant time with high probability for all operations in the word RAM model, e.g.~\cite{DH1990}. However, such dictionaries seem to require hash functions that are difficult to evaluate in parallel on the UWRAM. For instance,~\cite{DH1990} uses the modulo operator, for which we cannot see an obvious way to make a component-wise version. 

\paragraph*{Acknowledgments}
We would like to thank the anonymous reviewers of the conference version of this paper for their comments, which improved the presentation of the paper. In particular, we would like to thank the reviewer who suggested that it might be possible to strengthen the result by restricting the model to $w^{1+\epsilon}$-bit ultrawords.

\bibliographystyle{plainurl}
\bibliography{uwram_predecessor}

\begin{thebibliography}{10}

\bibitem{Ajtai1988}
Mikl{\'{o}}s Ajtai.
\newblock A lower bound for finding predecessors in {Y}ao's cell probe model.
\newblock {\em Comb.}, 8(3):235--247, 1988.
\newblock \href {https://doi.org/10.1007/BF02126797}
  {\path{doi:10.1007/BF02126797}}.

\bibitem{Andersson1996}
Arne Andersson.
\newblock Faster deterministic sorting and searching in linear space.
\newblock In {\em Proc. 37th {FOCS}}, pages 135--141, 1996.
\newblock \href {https://doi.org/10.1109/SFCS.1996.548472}
  {\path{doi:10.1109/SFCS.1996.548472}}.

\bibitem{AHNR1998}
Arne Andersson, Torben Hagerup, Stefan Nilsson, and Rajeev Raman.
\newblock Sorting in linear time?
\newblock {\em J. Comput. Syst. Sci.}, 57(1):74--93, 1998.
\newblock \href {https://doi.org/10.1006/jcss.1998.1580}
  {\path{doi:10.1006/jcss.1998.1580}}.

\bibitem{AFGV1997}
Lars Arge, Paolo Ferragina, Roberto Grossi, and Jeffrey~Scott Vitter.
\newblock On sorting strings in external memory (extended abstract).
\newblock In {\em Proc. 29th {STOC}}, pages 540--548, 1997.
\newblock \href {https://doi.org/10.1145/258533.258647}
  {\path{doi:10.1145/258533.258647}}.

\bibitem{BF2002}
Paul Beame and Faith~E. Fich.
\newblock Optimal bounds for the predecessor problem and related problems.
\newblock {\em J. Comput. Syst. Sci.}, 65(1):38--72, 2002.
\newblock \href {https://doi.org/10.1006/jcss.2002.1822}
  {\path{doi:10.1006/jcss.2002.1822}}.

\bibitem{Belazzougui2012}
Djamal Belazzougui.
\newblock Worst-case efficient single and multiple string matching on packed
  texts in the word-{RAM} model.
\newblock {\em J. Discrete Algorithms}, 14:91--106, 2012.
\newblock \href {https://doi.org/10.1016/j.jda.2011.12.011}
  {\path{doi:10.1016/j.jda.2011.12.011}}.

\bibitem{BBPV2009}
Djamal Belazzougui, Paolo Boldi, Rasmus Pagh, and Sebastiano Vigna.
\newblock Monotone minimal perfect hashing: searching a sorted table with
  {$O(1)$} accesses.
\newblock In {\em Proc. 20th {SODA}}, pages 785--794, 2009.
\newblock URL: \url{http://dl.acm.org/citation.cfm?id=1496770.1496856}.

\bibitem{BBV2010}
Djamal Belazzougui, Paolo Boldi, and Sebastiano Vigna.
\newblock Dynamic z-fast tries.
\newblock In {\em Proc. 17th {SPIRE}}, pages 159--172, 2010.
\newblock \href {https://doi.org/10.1007/978-3-642-16321-0\_15}
  {\path{doi:10.1007/978-3-642-16321-0\_15}}.

\bibitem{BFK2006}
Michael~A. Bender, Martin Farach{-}Colton, and Bradley~C. Kuszmaul.
\newblock Cache-oblivious string {B}-trees.
\newblock In {\em Proc. 25th {PODS}}, pages 233--242, 2006.
\newblock \href {https://doi.org/10.1145/1142351.1142385}
  {\path{doi:10.1145/1142351.1142385}}.

\bibitem{BEGV2018}
Philip Bille, Mikko~Berggren Ettienne, Inge~Li G{\o}rtz, and Hjalte~Wedel
  Vildh{\o}j.
\newblock Time-space trade-offs for {L}empel-{Z}iv compressed indexing.
\newblock {\em Theor. Comput. Sci.}, 713:66--77, 2018.
\newblock \href {https://doi.org/10.1016/j.tcs.2017.12.021}
  {\path{doi:10.1016/j.tcs.2017.12.021}}.

\bibitem{BGS2017}
Philip Bille, Inge~Li G{\o}rtz, and Frederik~Rye Skjoldjensen.
\newblock Deterministic indexing for packed strings.
\newblock In {\em Proc. 28th {CPM}}, pages 6:1--6:11, 2017.
\newblock \href {https://doi.org/10.4230/LIPIcs.CPM.2017.6}
  {\path{doi:10.4230/LIPIcs.CPM.2017.6}}.

\bibitem{BGS2022a}
Philip Bille, Inge~Li G{\o}rtz, and Frederik~Rye Skjoldjensen.
\newblock Partial sums on the ultra-wide word {RAM}.
\newblock {\em Theor. Comput. Sci.}, 905:99--105, 2022.
\newblock Announced at TAMC 2020.
\newblock \href {https://doi.org/10.1016/j.tcs.2022.01.002}
  {\path{doi:10.1016/j.tcs.2022.01.002}}.

\bibitem{BGS2022b}
Philip Bille, Inge~Li G{\o}rtz, and Tord Stordalen.
\newblock Predecessor on the ultra-wide word {RAM}.
\newblock In {\em Proc. 18th {SWAT}}, pages 18:1--18:15, 2022.
\newblock \href {https://doi.org/10.4230/LIPIcs.SWAT.2022.18}
  {\path{doi:10.4230/LIPIcs.SWAT.2022.18}}.

\bibitem{BLRS+2015}
Philip Bille, Gad~M. Landau, Rajeev Raman, Kunihiko Sadakane, Srinivasa~Rao
  Satti, and Oren Weimann.
\newblock Random access to grammar-compressed strings and trees.
\newblock {\em {SIAM} J. Comput.}, 44(3):513--539, 2015.
\newblock \href {https://doi.org/10.1137/130936889}
  {\path{doi:10.1137/130936889}}.

\bibitem{BCFKM2005}
Andrej Brodnik, Svante Carlsson, Michael~L. Fredman, Johan Karlsson, and J.~Ian
  Munro.
\newblock Worst case constant time priority queue.
\newblock {\em J. Syst. Softw.}, 78(3):249--256, 2005.
\newblock \href {https://doi.org/10.1016/j.jss.2004.09.002}
  {\path{doi:10.1016/j.jss.2004.09.002}}.

\bibitem{CRDI2007}
Thomas Chen, Ram Raghavan, Jason~N. Dale, and Eiji Iwata.
\newblock Cell {B}roadband engine architecture and its first implementation -
  {A} performance view.
\newblock {\em {IBM} J. Res. Dev.}, 51(5):559--572, 2007.
\newblock \href {https://doi.org/10.1147/rd.515.0559}
  {\path{doi:10.1147/rd.515.0559}}.

\bibitem{Intel2011}
Intel Corporation.
\newblock {Intel\textregistered} advanced vector extensions programming
  reference.
\newblock {\em Intel Corporation}, 2011.

\bibitem{DH1990}
Martin Dietzfelbinger and Friedhelm~Meyer auf~der Heide.
\newblock A new universal class of hash functions and dynamic hashing in real
  time.
\newblock In {\em Proc. 17th {ICALP}}, pages 6--19, 1990.
\newblock \href {https://doi.org/10.1007/BFb0032018}
  {\path{doi:10.1007/BFb0032018}}.

\bibitem{DHKP1997}
Martin Dietzfelbinger, Torben Hagerup, Jyrki Katajainen, and Martti Penttonen.
\newblock A reliable randomized algorithm for the closest-pair problem.
\newblock {\em J. Algorithms}, 25(1):19--51, 1997.
\newblock \href {https://doi.org/10.1006/jagm.1997.0873}
  {\path{doi:10.1006/jagm.1997.0873}}.

\bibitem{DKMHRT1994}
Martin Dietzfelbinger, Anna~R. Karlin, Kurt Mehlhorn, Friedhelm~Meyer auf~der
  Heide, Hans Rohnert, and Robert~Endre Tarjan.
\newblock Dynamic perfect hashing: Upper and lower bounds.
\newblock {\em {SIAM} J. Comput.}, 23(4):738--761, 1994.
\newblock \href {https://doi.org/10.1137/S0097539791194094}
  {\path{doi:10.1137/S0097539791194094}}.

\bibitem{Farach1997}
Martin Farach.
\newblock Optimal suffix tree construction with large alphabets.
\newblock In {\em Proc. 38th {FOCS}}, pages 137--143, 1997.
\newblock \href {https://doi.org/10.1109/SFCS.1997.646102}
  {\path{doi:10.1109/SFCS.1997.646102}}.

\bibitem{FLNS2015}
Arash Farzan, Alejandro L{\'{o}}pez{-}Ortiz, Patrick~K. Nicholson, and
  Alejandro Salinger.
\newblock Algorithms in the ultra-wide word model.
\newblock In {\em Proc. 12th {TAMC}}, pages 335--346, 2015.
\newblock \href {https://doi.org/10.1007/978-3-319-17142-5\_29}
  {\path{doi:10.1007/978-3-319-17142-5\_29}}.

\bibitem{FKS1984}
Michael~L. Fredman, J\'{a}nos Koml\'{o}s, and Endre Szemer{\'e}di.
\newblock Storing a sparse table with {$O(1)$} worst case access time.
\newblock {\em J. ACM}, 31(3):538--544, 1984.
\newblock \href {https://doi.org/10.1145/828.1884}
  {\path{doi:10.1145/828.1884}}.

\bibitem{FW1993}
Michael~L. Fredman and Dan~E. Willard.
\newblock Surpassing the information theoretic bound with fusion trees.
\newblock {\em J. Comput. Syst. Sci.}, 47(3):424--436, 1993.
\newblock \href {https://doi.org/10.1016/0022-0000(93)90040-4}
  {\path{doi:10.1016/0022-0000(93)90040-4}}.

\bibitem{Hagerup1998}
Torben Hagerup.
\newblock Sorting and searching on the word {RAM}.
\newblock In {\em Proc. 15th {STACS}}, pages 366--398, 1998.
\newblock \href {https://doi.org/10.1007/BFb0028575}
  {\path{doi:10.1007/BFb0028575}}.

\bibitem{Han2004}
Yijie Han.
\newblock Deterministic sorting in {$O(n\log\log n)$} time and linear space.
\newblock {\em J. Algorithms}, 50(1):96--105, 2004.
\newblock \href {https://doi.org/10.1016/j.jalgor.2003.09.001}
  {\path{doi:10.1016/j.jalgor.2003.09.001}}.

\bibitem{LP2012}
Kasper~Green Larsen and Rasmus Pagh.
\newblock {I/O}-efficient data structures for colored range and prefix
  reporting.
\newblock In {\em Proc. 23rd SODA}, pages 583--592, 2012.
\newblock \href {https://doi.org/10.1137/1.9781611973099.49}
  {\path{doi:10.1137/1.9781611973099.49}}.

\bibitem{LMSTBK1999}
R.~Leben, M.~Miletic, M.~\u{S}pegel, A.~Torst, A.~Brodnik, and K.~Karlsson.
\newblock Design of high performance memory module on {PC100}.
\newblock In {\em Proc. Electrotechnical and Computer Science Conference
  {(ERK)}}, pages 75--78, 1999.

\bibitem{Miltersen1994}
Peter~Bro Miltersen.
\newblock Lower bounds for union-split-find related problems on random access
  machines.
\newblock In {\em Proc. 26th {STOC}}, pages 625--634, 1994.
\newblock \href {https://doi.org/10.1145/195058.195415}
  {\path{doi:10.1145/195058.195415}}.

\bibitem{MNSW1998}
Peter~Bro Miltersen, Noam Nisan, Shmuel Safra, and Avi Wigderson.
\newblock On data structures and asymmetric communication complexity.
\newblock {\em J. Comput. Syst. Sci.}, 57(1):37--49, 1998.
\newblock \href {https://doi.org/10.1006/jcss.1998.1577}
  {\path{doi:10.1006/jcss.1998.1577}}.

\bibitem{NR2020}
Gonzalo Navarro and Javiel Rojas{-}Ledesma.
\newblock Predecessor search.
\newblock {\em {ACM} Comput. Surv.}, 53(5):105:1--105:35, 2020.
\newblock \href {https://doi.org/10.1145/3409371} {\path{doi:10.1145/3409371}}.

\bibitem{PT2006}
Mihai P{\u{a}}tra{\c{s}}cu and Mikkel Thorup.
\newblock Time-space trade-offs for predecessor search.
\newblock In {\em Proc. 38th {STOC}}, pages 232--240, 2006.
\newblock \href {https://doi.org/10.1145/1132516.1132551}
  {\path{doi:10.1145/1132516.1132551}}.

\bibitem{PT2007}
Mihai P{\u{a}}tra{\c{s}}cu and Mikkel Thorup.
\newblock Randomization does not help searching predecessors.
\newblock In {\em Proc. 18th {SODA}}, pages 555--564, 2007.
\newblock URL: \url{http://dl.acm.org/citation.cfm?id=1283383.1283443}.

\bibitem{PT2014}
Mihai P{\u{a}}tra{\c{s}}cu and Mikkel Thorup.
\newblock Dynamic integer sets with optimal rank, select, and predecessor
  search.
\newblock In {\em Proc. 55th {FOCS}}, pages 166--175, 2014.
\newblock \href {https://doi.org/10.1109/FOCS.2014.26}
  {\path{doi:10.1109/FOCS.2014.26}}.

\bibitem{Reinders2013}
James Reinders.
\newblock {Intel\textregistered} {AVX-512} instructions.
\newblock {Intel\textregistered} Corporation, 2013.

\bibitem{SV2008}
Pranab Sen and Srinivasan Venkatesh.
\newblock Lower bounds for predecessor searching in the cell probe model.
\newblock {\em J. Comput. Syst. Sci.}, 74(3):364--385, 2008.
\newblock \href {https://doi.org/10.1016/j.jcss.2007.06.016}
  {\path{doi:10.1016/j.jcss.2007.06.016}}.

\bibitem{SBBE+2017}
Nigel Stephens, Stuart Biles, Matthias Boettcher, Jacob Eapen, Mbou Eyole,
  Giacomo Gabrielli, Matt Horsnell, Grigorios Magklis, Alejandro Martinez,
  Nathana{\"{e}}l Pr{\'{e}}millieu, Alastair Reid, Alejandro Rico, and Paul
  Walker.
\newblock The {ARM} scalable vector extension.
\newblock {\em {IEEE} Micro}, 37(2):26--39, 2017.
\newblock \href {https://doi.org/10.1109/MM.2017.35}
  {\path{doi:10.1109/MM.2017.35}}.

\bibitem{Boas1977}
Peter van Emde~Boas.
\newblock Preserving order in a forest in less than logarithmic time and linear
  space.
\newblock {\em Inform. Process. Lett.}, 6(3):80--82, 1977.
\newblock \href {https://doi.org/10.1016/0020-0190(77)90031-X}
  {\path{doi:10.1016/0020-0190(77)90031-X}}.

\bibitem{BKZ1997}
Peter van Emde~Boas, R.~Kaas, and E.~Zijlstra.
\newblock Design and implementation of an efficient priority queue.
\newblock {\em Math. Syst. Theory}, 10:99--127, 1977.
\newblock \href {https://doi.org/10.1007/BF01683268}
  {\path{doi:10.1007/BF01683268}}.

\bibitem{Willard1983}
Dan~E. Willard.
\newblock Log-logarithmic worst-case range queries are possible in space
  {$\Theta(N)$}.
\newblock {\em Inform. Process. Lett.}, 17(2):81--84, 1983.
\newblock \href {https://doi.org/10.1016/0020-0190(83)90075-3}
  {\path{doi:10.1016/0020-0190(83)90075-3}}.

\end{thebibliography}

\appendix

\section{Blend and \texorpdfstring{$\boldsymbol{2w}$}{2w}-bit Multiplication}\label{sec:2wbitmult}
\subsection{Supporting Blend} 
{Given the ultrawords \uw{X}, \uw{Y} and \uw{I} where each component of \uw{I} is either $0$ or $1$, we define the \emph{componentwise blend} of $X$, $Y$, and $I$ to be the ultraword $Z$ such that $Z\angle{i} = X\angle{i}$ if $I\angle{i} = 0$ and $Y\angle{i}$ if $I\angle{i} = 1$. To compute the blend in constant time we do as follows. Compute $\uw{I'} = \angle{0,\ldots,0} - I$; then $\uwi{I'}{i}$ contains only $1$-bits if $\uwi{I}{i} = 1$ and only $0$-bits otherwise, since $0 - 1 \mod 2^w = 2^w - 1$. Then the blend of $X$ and $Y$ can be computed by $(X~\& ~ \sim I') ~\mid ~(Y~\& ~I')$.

\subsection{Supporting \texorpdfstring{$\boldsymbol{2w}$}{2w}-Bit Componentwise Multiplication}
\begin{figure}
    \centering
    \includegraphics[scale=0.55]{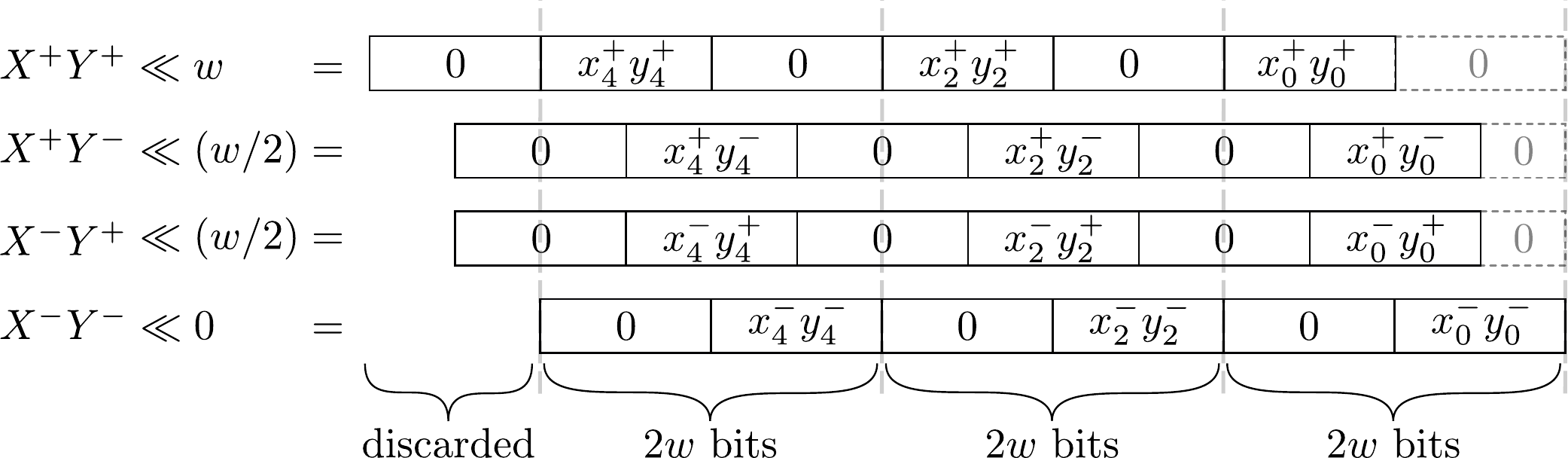}
    \caption{Illustrates step~3 of $2w$-bit multiplication. Each of the products \uw{X^+Y^+}, \uw{X^+Y^-}, \uw{X^-Y^+} and \uw{X^-Y^-} are left-shifted by respectively $w$, $w/2$, $w/2$ and $0$ by shifting in zeroes from the right. Then they are added together using componentwise addition for $2w$-bit components. Since what we sum up in a $2w$-bit component adds up to the product of two $w$-bit integers, we only need $2w$ bits to store the result. Hence the addition will not overflow.}
    \label{fig:2w_bit_multiplication_step_3}
\end{figure}

We show how to implement $2w$-bit componentwise multiplication in constant time.  
Let $x^+$ and $x^-$ denote the leftmost and rightmost half of the binary representation of $x$, respectively. Then, if $x$ is a $2k$-bit integer we have that $x = x^+2^k + x^-$ where $x^+ = x/2^k$ and $x^- = x \mod 2^k$. Given $\uw{X} = \langle 0 , x_{w - 2}, \ldots,0, x_2, 0, x_0 \rangle$ and $\uw{Y} = \langle 0 , y_{w - 2}, \ldots, 0, y_2, 0, y_0 \rangle$, recall that the $2w$-bit componentwise multiplication of \uw{X} and \uw{Y} is the ultraword $\uw{Z} = \langle z_{w-2}^+,z_{w-2}^-, \ldots,z_2^+,z_2^-,z_0^+,z_0^-\rangle$ where $z_i$ is the $2w$-bit product of $x_i$ and $y_i$. 

The main idea for computing \uw{Z} is to use the identity
\begin{equation}
    \label{eq:multiplication_decomposition}
        \begin{aligned}
             xy &= (x^+2^{w/2} + x^-)(y^+2^{w/2} + y^-)\\
                &= x^+y^+2^w + (x^+y^- + x^-y^+)2^{w/2} + x^-y^-
    \end{aligned}
\end{equation}
where $x$ and $y$ are $w$-bit integers. We simulate this in parallel as follows.\\

\paragraph*{Step 1: Compute $x_i^+$, $x_i^-$, $y_i^+$ and $y_i^-$ for all even $i$.} We first construct \uw{X^+} and \uw{X^-} such that $\uwi{X^+}{i} = x_i^+$ and $\uwi{X^-}{i} = x_i^-$ for even $i$ and zero otherwise, and similarly for \uw{Y}. Compute the integer $m = 2^{w/2} - 1$ which consists of $w/2$ zeroes followed by $w/2$ ones. Load $m$ into \uw{M}. Compute $\uw{X^-} = \uw{X}~\&~\uw{M}$ and $\uw{X^+} = (\uw{X}\gg w/2)~\&~\uw{M}$. Compute \uw{Y^+} and \uw{Y^-} in the same way.\\

\paragraph*{Step 2: Compute the products of the $w/2$-bit integers.}
Use componentwise multiplication to compute each of the ultrawords $\uw{X^+Y^+}, \: \uw{X^+Y^-}, \: \uw{X^-Y^+} \: \text{and} \: \uw{X^-Y^-}$. Since each component of $X^+$, $X^-$, $Y^+$ and $Y^-$ is a $(w/2)$-bit integer, no overflow occurs. The odd components still store $0$.

\paragraph*{Step 3: Align and add the products.}
Align the products by left-shifting them the amount specified in Equation~\ref{eq:multiplication_decomposition}, i.e.
\[ \uw{X^+Y^+} \ll w \qquad \uw{X^+Y^-}\ll w/2 \qquad \uw{X^-Y^+} \ll w/2 \qquad \uw{X^-Y^-} \ll 0  \]
Add the aligned ultrawords using componentwise addition for $2w$-bit components (see e.g. Hagerup~\cite{Hagerup1998}) and return the result. See Figure~\ref{fig:2w_bit_multiplication_step_3} for an illustration. Since the sum of the terms added together in a $2w$-bit component exactly correspond to the multiplication of two $w$-bit integers, the addition will not overflow.\\

Bitwise \&, left- and right-shifts, componentwise multiplication and componentwise additions for arbitrary component sizes all run in constant time. Each step uses a constant number of these operations, so the procedure runs in constant time. 

\end{document}